\documentclass[tighten,twocolumn]{elsart3p}

\usepackage{amssymb,graphicx,epsfig}

\begin{document}

\title{Constraint preserving boundary conditions for the  Ideal Newtonian MHD equations}

\author{Mariana C\'ecere${}^1$, Luis Lehner${}^2$ and Oscar Reula${}^1$ }
\address{${}^1$ FaMAF, Universidad Nacional de C\'ordoba, C\'ordoba, 5000 (Argentina)\\
${}^2$ Department of Physics and Astronomy, Louisiana State
University, Baton Rouge, LA 70803-4001}

\begin{abstract}
We study and develop constraint preserving boundary
conditions for the Newtonian magnetohydrodynamic equations and analyze
the behavior of the numerical solution upon considering different
possible options.
\end{abstract}

\maketitle


\section{Introduction}
Magnetic fields play an important role in the behavior
of plasmas and are thought mediate important effects like dynamos
in the core of planets and the formation of jets in active galactic nuclei
and gamma ray bursts; induce a variety of magnetic instabilities;
realize solar flares, etc. (see e.g. \cite{book1,book2}).
Understanding the role of magnetic fields
in these and other phenomena have spurred through the years
many efforts to obtain solutions of the magnetohydrodynamic
equations. The non-linear nature of these equations limits the understanding
that can be gained in a particular problem via analytical techniques. This implies that
solutions for complex systems must be obtained by numerical means and a suitable
numerical implementation must be constructed for this purpose.
Such implementation must be able to evolve the solution to the future of
some initial configuration and guarantee its quality. A delicate, subsidiary
quantity,  can be monitored in part to estimate this. This quantity is the ``monopole
constraint'' $\partial_i B^i$ which must be zero at the analytical level for
a consistent solution. This quantity is not a part of the main variables,
rather it is a derived quantity which should be satisfied by a true physical
solution. In practice, unless a numerical implementation  of the MHD equations is
carefully designed this constraint can be severely violated. This, in turn,
signals (and is sometimes the cause) of a degrading numerical solution.
For these reasons, several approaches have been investigated and developed for
guaranteeing a controlled behavior of this quantity.
One such approach is known as the {\it constraint transport technique}~\cite{constrainttransport,toth}
which adopts a particular algorithm that staggers the variables appropriately
to ensure the satisfaction of the constraint at round-off level within Finite Difference and
Finite Elements techniques. This approach has been
quite successful in a number of applications across different disciplines and particularly
relevant in astrophysics
applications~\cite{Gammie:2003rj,MHD_stone,MHD_hawley,MHD_shibata,MHD_shapiro,MHD_rezzolla,Li:2006jj,Obergaulinger:2005xd}.
However,
by design it imposes limits on the algorithmic options available to an implementation.
This fact can be at odds, or introduce complications, with applications where adaptive mesh refinement
is required and/or advanced numerical techniques (that exploit useful
properties of the equations) are adopted. An alternative approach, which controls
the constraint at truncation-error level maintains complete freedom in the
numerical techniques to be adopted. This approach, referred to as
{\it divergence cleaning} puts the burden to control the constraint not
on the algorithm to be employed but rather on the system of equations to be
solved itself \cite{kimclean,dedner_2002}. This is achieved by considering an additional variable
suitably coupled to the system through another equation so that, through the evolution,
the constraint behavior is kept under control. While this method has, to date, received less attention
than the constraint transport one, successful applications in diverse scenarios have already
illustrated its usefulness (e.g. \cite{dedner_2002,balsarakim,toth,neilsen1,neilsen2}).

Regardless of the technique employed, boundary conditions can play a
significant role and may spoil all efforts towards a correct implementation
in the absence of boundaries. Clearly, even when a stable option is adopted, it need not be
consistent with the goal of preserving the constraint (be it at round-off
or truncation levels) and so carefully designed boundary conditions must
be formulated. Commonly employed options include straightforward outflow-type
conditions --which do not enforce the constraint-- or ``absorbing'' boundary conditions
which aim to reduce the influence of spurious effects induced at the boundary in the numerical
solution~\cite{dednerbdry}.

In the present work we concentrate on formulating constraint preserving
boundary conditions for the Newtonian ideal MHD equations (for systems with and without
divergence cleaning). Such a task
is intimately related to the hyperbolic properties of the equations and
so we re-analyze the system of equations and discuss alternatives for controlling
the constraints through the divergence cleaning technique.

We therefore organize our presentation along the following lines. In section II
we analyze the system of equations and the formulation of boundary conditions
beginning with a simplified model from which we draw the strategy to apply in the
complete MHD system. Section III presents a series of tests that highlight the
benefits gained by our construction. We conclude in section IV with some final
comments.

\section{The Newtonian equations \& constraint hyperbolic cleaning}
The equations describing the ideal Newtonian MHD equations in terms of
the variables $(\rho, p, v^i, B^i)$ are \cite{book1,book2}):
\begin{eqnarray}
\label{eqn:NMHD}
\partial_t \rho     \ &=& \ -\nabla_i(\rho v^i) \, ,  \nonumber \\
\partial_t p        \ &=& \ - v^i \nabla_i p - \gamma p \nabla_jv^j \, , \nonumber \\
\rho \partial_t v^i \ &=& \ -\rho v^j \nabla_j v^i - \nabla^i p - \nabla_j( e^{ij}\frac12 B_kB^k - B^j B^i) \, , \nonumber \\
\partial_t B^i      \ &=& \ -\nabla_j(v^j B^i - v^i B^j) \, .\nonumber
\end{eqnarray}

In this form, this system of equations is {\it weakly hyperbolic} since there is
no complete set of eigenvectors. This indicates instabilities are likely to arise
unless some modes are carefully controlled. The constraint transport technique attempts
to do so at the algorithmical level by enforcing $\partial_i B^i=0$ at the discrete level.
We here choose an alternative approach where the situation is remedied at the analytical level through
the addition of an extra field coupled to the system in a suitable way \cite{kimclean,dedner_2002}.
To simplify the discussion, we first consider a simpler model which captures essential features
of the main system.

\subsection{Preliminaries  \label{boundarytoy}}
To illustrate this approach consider first the system obtained when
the fluid field variables as given and stationary.
The resulting system describes a simple model that shares key problematic features 
of the original ideal MHD system (\ref{eqn:NMHD}),
although in the latter case additional aspects enter under consideration. In this simple
case one just has the magnetic field which evolves under,

\begin{equation}
\label{eqn:B_alone}
\partial_t B_i \ = \ -\nabla_j(v^j B^i - v^i B^j) \, .
\end{equation}
This equation is already weakly hyperbolic, since a plane-wave analysis indicates that
when the wave number vector is perpendicular to the velocity vector there exists only two linearly independent eigenvectors.
To see this consider the case where the wave number vector is normal to the velocity vector, and define 
coordinates axis such that the first axis lies along the wave vector --so $k_i=(k,0,0)$-- the second along 
the velocity vector, --so $v^i=(0,v,0)$-- and the third perpendicular to 
both of them. Then the eigenvalue-eigenvector problem, obtained by expressing 
$B^i = U e^{\sigma t + k_i x^i}$, gives rise to the following problem:
$\sigma U = M U$ (with $\sigma$ and $U$ a frequency and a vector to be determined) and $M$ defined as

\begin{equation}
 M = \left( \begin{array}{ccc}
         0 & 0 & 0 \\
         kv & 0 & 0 \\
         0 & 0 & 0
        \end{array}
\right)
\end{equation}
which is clearly non diagonalizable. Thus, there is no stability in the usual $L^2$ norm sense.
%
%
Interestingly the divergence of the magnetic field is preserved by equation (\ref{eqn:B_alone}),
that is, if one defines $D:=\nabla_kB^k$ and takes the divergence of (\ref{eqn:B_alone})
one obtains the {\em induced evolution equation} for $D$ as $\partial_t D=0$.
Hence,  if $D$ is zero initially it remains so as along as the time integration lines
do not intersect a boundary. However, we note that if instead of considering
the $L_2$ norm, one considers a different norm defined by adding
 a term proportional to the divergence of
$B_i$, namely a norm of the type:

\[
 {\cal{E}}:= \int_{\Sigma_t} [B^2 + c (\nabla_kB^k)^2] dV
\]
one can show that $\dot {\cal{E}}=0$ and so this energy is controlled.
Therefore, from an analytical point of view equation (\ref{eqn:B_alone}) is not a bad one.

At the numerical level, things become more delicate as  generic
violations of this constraint will arise due to truncation or round-off errors which
might grow unless a careful implementation of the equations is adopted.
Furthermore, even when an integration scheme is available that controls the
constraint in the absence of boundaries,
(which is not difficult to obtain in flat space in Cartesian coordinates
--which we use in this work--), the boundary conditions modify the equations at the boundaries
which can cause the constraint to grow and propagate  to the interior. Thus, we next discuss
a way to achieve a more robust behavior. Rather than doing so by the particular
numerical algorithm employed we work first at the analytical level and adopt a strategy
that would help even beyond the simple problem adopted here.

To remedy the lack of strong hyperbolicity and good constraint
propagation a possible approach can be adopted which makes use of the freedom to
add to equation (\ref{eqn:B_alone}) any term
proportional to that divergence. The proposed system is (in analogy with that in \cite{dedner_2002}):

\begin{eqnarray}
\label{eqn:B_phi}
\partial_t B_i  \ &=& \ -\nabla_j(v^j B^i - v^i B^j) - c_l \nabla_i \phi \, , \nonumber \\
\partial_t \phi \ &=& \ -c_l \nabla_k B^k - s \phi \, .
\end{eqnarray}

Notice that for $\phi=0$ the modified system is equivalent to the original one,
and $\phi$ has as a source proportional to the constraint. Thus if boundary and initial data
are such as to guarantee that $\nabla_kB^k=0$, and trivial
data is given for $\phi$ the solution to the modified system is also a solution of the original equation.
The advantage of this
modification is that now the system is strongly hyperbolic, so it has a well posed initial problem and is
stable irrespective of whether or not $\nabla_kB^k$ vanishes. Furthermore, the induced
evolution equation for $D$ is no
longer trivial, and it implies that the field $D$ propagates with speed $c_l$.
Consequently one
can make the constraint violations to propagate away from the integration region, and, if correct
boundary conditions are given, leave the computational region entirely. Additionally,
the last term in the equation for $\phi$ in (\ref{eqn:B_phi}) induces a decay of the constraint
 for $s>0$ as it travels along the
integration region further helping to keep it under control.

Drawing from this exercise, other modifications are certainly possible; for instance,
 one could consider a more complex system given by,

\begin{eqnarray}
\partial_t B_i  \ &=& \ - u^j\nabla_j B_i + B^j \nabla_j u_i - B_i \nabla_j u^j + (1-\alpha) u^i \nabla_j B^j - c_l \nabla_i \phi \nonumber \\
\partial_t \phi \ &=& \ - \beta u^j\nabla_j \phi - c_l \nabla^i B_i - s \phi \, .
\end{eqnarray}
Which, for a range of values of the parameters $\alpha$ and $\beta$ for is strongly hyperbolic.  Among
these, of particular interest is the system with $\alpha=\beta=1$, which is strongly  hyperbolic even in
the limit $c_l \to 0$, so its associated initial value problem would
be well posed even without the coupling to the field $\phi$. Another
interesting choice is the one
$\alpha=\beta=0$ since the resulting system can be expressed in  conservative form.
This property is often preferred in applications where shocks are present as one can easily take
advantage of special algorithms defined to deal
 with these issues (see e.g. \cite{leveque,zachary,balsaraM,dezeu,hanoi}).

\subsection*{Boundary conditions}
To fix ideas, we now study the possible boundary conditions for the above system (with $\alpha=\beta=0$). Our
goal is to define conditions which can yield a well posed problem and, if possible, guarantee no
violations are introduced into the computational domain. To this end we must determine which are the
incoming and outgoing modes off the boundary surface as this information is key to understand
what data is freely specifiable. For this we seek a solution of the form
$(B^i,\phi) = U e^{\sigma t + n_i x^i}$, where $U$ is a vector to be determined along with the frequency
$\sigma$, and $n_i$ is the outgoing normal to the boundary under consideration. Applying this
solution to the system we obtain:

\begin{eqnarray}
 \sigma B^i  \ &=& \ - B^i v_n + v^i B_n - c_l n^i \phi \, ,\nonumber \\
 \sigma \phi \ &=& \ - c_l B_n \, ,
\end{eqnarray}
where $B_n:= B^in_i$, and $v_n:= v^in_i$. This problem of
eigenvalues/eigenvectors can be solved to determine the incoming ($\sigma
> 0$), tangential ($\sigma=0$) and outgoing  ($\sigma < 0$) modes. A necessary condition for
a well posed problem for hyperbolic systems indicates that data must
be given only to the incoming modes, since the others are determined
from the inside of the integration region~\cite{reula_2006_a,GKO}. Recall that the incoming
modes can be defined even as linear functions of the outgoing ones
so long as the coefficients are small enough.

At a given boundary, there are three eigenvalues, $(\sigma_0 := -v_n, \sigma_{\pm}:= \pm c_l)$. The first
corresponds to two linearly independent modes which span the tangent space of the boundary point (i.e.
they are perpendicular to $n_i$). They are positive whenever the velocity is incoming, these are the
modes which are dragged along $v^i$. The other eigenvalues correspond to the cases where we can choose
$B_n:=b$ arbitrarily and,
\begin{eqnarray}
\phi &=& \mp b \, \, ,\nonumber \\
B^i &=& \frac{(v^i \pm n^i c_l)b}{v_n \pm c_l} \, .\nonumber
\end{eqnarray}
Notice one can always adopt a value for $c_l$ larger than the maximum velocity expected so that
denominator is never zero. The eigenvector corresponding to the positive eigenvalue
determine the combination of variable that must be always suitably given to preserve
the constraint. The complete expression for the
eigenbase and its co-base is.

\begin{eqnarray}
 U^{0}_{1} &:=& (0,e_1,0) \;\;\; , \;\;\; \sigma = -v_n \; ,\nonumber \\
 U^{0}_{2} &:=& (0,e_2,0) \;\;\; , \;\;\; \sigma = -v_n \; ,\nonumber \\
 U^{\pm}   &:=& (1,\frac{(v^i \pm n^i c_l)}{v_n \pm c_l},\mp1) \;\;\; , \;\;\; \sigma =\pm c_l\nonumber \, ,
\end{eqnarray}
where the generic vector is $U:=(B_n,\tilde{B}^i,\phi)$ with $\tilde{B}^in_i=0$, and $\{e_1,e_2\}$ are two orthonormal vectors on the tangent of the boundary.

The co-base is given by
\begin{eqnarray}
 \theta_{0}^{1} &:=& (\frac{v_n v_1}{c_l^2-v_n^2},e_1,\frac{c_l v_1}{c_l^2-v_n^2}) \, ,\nonumber \\
 \theta_{0}^{2} &:=& (\frac{v_n v_2}{c_l^2-v_n^2},e_2,\frac{c_l v_2}{c_l^2-v_n^2}) \, ,\nonumber \\
 \theta_{\pm} &:=& \frac12 (1,0,\mp1) \, , \nonumber
\end{eqnarray}
where $v_1:=e_1^iv_i$, and $v_2:=e_2^iv_i$.
In order to see how we must define boundary values consistent with the constraints
we now analyze the subsidiary system determining the evolution of the constraint.
In order to treat it as a first order system we define a new variable $\delta_i := \nabla_i \phi$,
whose evolution equation is determined by (the time derivative of) the evolution equation of $\phi$.
The full subsidiary system is then:

\begin{eqnarray}
\partial_t D \ &=& \ -c_l \nabla^i \delta_i \, , \nonumber \\
\partial_t \delta_i \ &=& \ -c_l \nabla_i D - s \delta_i \, .
\end{eqnarray}
Here again we must investigate now which are the incoming and outgoing modes for this system. If we can
impose boundary conditions to the main system so that we ensure no incoming mode is created for
this subsidiary system, then uniqueness of the (trivial) solution would guarantee that the solution
of the
main system has vanishing constraint quantities and so it is a solution of the original system.
As done previously,
plane wave solutions of the above system perpendicular to the boundary give rise to the following
eigenvalue/eigenvector problem:

\begin{eqnarray}
\kappa D \ &=& \ -c_l n^i \delta_i \, , \nonumber \\
\kappa \delta_i \ &=& \ -c_l n_i D \, ,
\end{eqnarray}
and its eigenvalues are $\{\kappa_0=0,\kappa_{\pm}=\pm c_l\}$ with the corresponding eigenvectors:

\begin{eqnarray}
 V^{0}_{1} &:=& (0,e_1,0) \;\;\; , \;\;\; \kappa_0 = 0 \, , \nonumber \\
 V^{0}_{2} &:=& (0,e_2,0) \;\;\; , \;\;\; \kappa_0 = 0 \, , \nonumber \\
 V^{\pm} &:=& (1,\frac{(v^i \pm n^i c_l)}{v_n \pm c_l},\mp 1) \;\;\; , \;\;\; \kappa_{\pm}=\pm c_l \,  ,\nonumber
\end{eqnarray}
where the generic vector is $V:=(\delta_n,\tilde{\delta}^i,D)$ with $\tilde{\delta}^in_i=0$, and
$\{e_1,e_2\}$ are two orthonormal vectors tangent to the boundary.

The co-base is given by
\begin{eqnarray}
 \Theta_{0}^{1} &:=& (0,e_1,0) \, , \nonumber \\
 \Theta_{0}^{2} &:=& (0,e_2,0) \, , \nonumber \\
 \Theta_{\pm} &:=& \frac12 (1,0,\mp1) \, .\nonumber
\end{eqnarray}
Here again the theory of boundary conditions indicate we should give as
a boundary condition the incoming mode as a linear function of the outgoing one.
That is we should ensure that the boundary data for the main system is such that:
\begin{equation}
 \Theta_+(V) + a \Theta_-(V) \hat{=} 0 \;\;\; , \;\;\; |a|<1
\end{equation}
This is a linear combination of space derivatives of the fields and can be implemented in
several different ways depending on the problem at hand. Here we choose to implement it at
the level of the evolution equations, namely by modifying the  evolution equations at the
boundary so that the constraint is satisfied there. From the evolution equations of the main
system we can solve for $D$ and $\delta_n$ in terms of the time derivatives of $B_n$ and
$\phi$ (and the remaining equations). Hence
\begin{eqnarray}
\delta_n &=& \frac{-1}{c_l}(\partial_t B_n - F_n) \, ,\nonumber \\
D &=& \frac{-1}{c_l}(\partial_t \phi- F_{\phi}) \, ,
\end{eqnarray}
with $F_n = -n_i\partial_j(u^jB^i-u^iB^j)$, and  $F_{\phi} = -s \phi$.\\
We can then translate the above boundary condition into:
\begin{equation}
- \dot B_n (1+a) + \dot \phi (1-a) + F_n (1+a) - F_{\phi} (1-a) \
\hat{=} \ 0 \, ,
\end{equation}
which can be re-expressed as:
\begin{equation}
-2 \left ( \theta_+(U) +  a \theta_-(U) \right ) + F_n (1+a) - F_{\phi} (1-a) \
\hat{=} \ 0 \, .
\end{equation}
For the particular case of setting the incoming constraint modes to zero ($a=0$) this implies,
\begin{equation}
 \theta_+(U) \ \hat{=} \ \frac{1}{2} \left( F_{\phi}
- F_n \right ) \label{eq:cpbc} \, .
\end{equation}
Thus, the boundary condition defined by the equation above
is not only maximally dissipative but also enforces the constraint.

\subsection{The complete system}
We now turn our attention to the complete system and develop an analogous strategy
following the discussion considered above. In most cases the literature
on the subject considers the MHD system expressed in terms variables
$U=(\rho,e,u^i,B^i)$ rather than $U=(\rho,p,u^i,B^i)$
--ie the internal energy $e$ instead of the pressure $p$--.
While this involves a simple change of variables it turns out the characteristic
decomposition is simpler in the latter case. So, our evolution system will be given
by the former, and we will obtain the characteristic structure for the latter followed by
a straightforward change of variables.  Our system of interest is a variation of the
initial MHD equations as (see \cite{dedner_2002}):

\begin{eqnarray}\label{mainsystem}
\partial_t \rho     \ &=& \ -\nabla_i(\rho v^i)  \, , \nonumber \\
\rho \partial_t v^i \ &=& \ -\rho v^j \nabla_j v^i - \nabla^i p -
B_k (\nabla^i B^k - \nabla^k B^i) -\alpha B^i \nabla_kB^k \, ,\nonumber \\
\partial_t B^i      \ &=& \ -\nabla_j(u^j B^i - u^i B^j) - \alpha u^i \nabla_jB^j - c_l \nabla^i \phi \, ,\nonumber \\
\partial_t e        \ &=& \ -\nabla_i((e+p+\frac12 B^2)v^i - B^i v\cdot B) -\alpha v^iB_i \nabla_k B^k -c_l B^k \nabla_k \phi \, ,\nonumber \\
\partial_t \phi     \ &=& \ - \alpha u^j \nabla_j \phi - c_l \nabla_jB^j -s \phi \, ,
\end{eqnarray}
where
\begin{equation}
p:= (\gamma-1)\left(e - \frac12 \rho v^2 - \frac12 B^2\right)
\;\;\; , \;\;\; c_s^2 := \frac{\gamma p}{\rho}
\end{equation}
Thys system includes both the possibility of divergence cleaning ($c_l\neq 0$) and 
Galilean invariance ($\alpha=1$) and is strongly hyperbolic, hence has
a complete set of eigenvectors. This will allow us to introduce a
new boundary treatment which ensures no constraint violations are introduced
through the computational domain boundaries. The parameter $\alpha$ controls the freedom of adding
the constraint equation. In what follows, we shall use the values $\alpha =0,1$.
The value $\alpha =0$ corresponds to the purely conservative system (notice that
velocity equation can be re-written with the help of the first equation as
momentum conservation). The  case with $\alpha =1$  gives rise to a system that
is Galilean invariant (see \cite{dedner_2002}) and also to the case where the
system is strongly hyperbolic \cite{godunov_1972,barth_1997} irrespective of the
field $\phi$. Hence one can study the  limit where the $\phi$ field decouples and
the constraint propagates along the velocity field.

\subsection*{Boundary system}
As done previously, we now consider a boundary with outgoing unit
normal $n^i$ and obtain the characteristic decomposition at this boundary.
For notational purposes we will employ an overbar to denote the perturbation
on a given variable, for instance $\bar \rho$ will denote the
perturbation of $\rho$ which will be considered as a fixed background quantity.
The characteristic decomposition is then determined from the system,
\begin{eqnarray}
\sigma \bar{\rho} \ &=& \ -v_n \bar{\rho} - \rho \bar{v_n} \, , \nonumber \\
\sigma \bar{v}_i \ &=& \ -v_n \bar{v}_i + B_i \bar{B}_n/\rho + (B_n/\rho) \bar{B}_i - n_i (B_j\bar{B}^j + \bar{p}) /\rho \, ,\nonumber \\
\sigma \bar{B}_i \ &=& \ -v_n \bar{B}_i + v_i\bar{B}_n - B_i\bar{v}_n + B_n\bar{v}_i +n_i \bar{\phi} \, , \nonumber \\
\sigma \bar{p} \ &=& \ -c_s^2\rho \bar{v}_n - (\gamma-1)B_jv^j\bar{B}_n - v_n \bar{p} +v(1-\gamma)B_n\bar{\phi} \, .\nonumber \\
\sigma \bar{\phi} \ &=& \ c_l^2 \bar{B}_n
\end{eqnarray}

The solution to this system is cumbersome and requires dealing with
different cases. The full analysis and solution is presented in appendix \ref{chardecomp}
and expressed in terms of a vector whose entries are: $U = (\bar{\rho}, \ \bar{v}_n, \
\tilde{\bar{v}}_i, \ \bar{B}_n, \ \tilde{\bar{B}}_i, \ \bar{p}, \
\bar{\phi})^T$ indicating perturbations off ($\rho, v^i n_i, \tilde v^i, B^i n_i, \tilde B^i, p, \phi)^T$
respectively with $\tilde v^i, \tilde B^i$ indicating the components of $v^i$ and $B^i$ orthogonal
to the boundary. While the full characteristic decomposition is lengthy, the characteristic
decomposition associated to $B_n$ and $\phi$ is particularly simple. In fact, the associated
co-basis is given by
\begin{equation}
\Theta_{L\pm} =\frac{1}{2 B_n} (0,0,0,1,0,0,\pm c_l^{-1}) \, ,
\end{equation}
i.e. involving essentially just $B_n$ and $\phi$, 
hence the imposition of suitable boundary conditions will be directly related to
that of our discussion of the simplified model in section \ref{boundarytoy}.

\subsection{Boundary conditions}
Having the complete characteristic structure it is now rather
straightforward to determine the possible boundary conditions.
A simple one is to set all incoming modes to zero but this,
generically, will not be consistent with the constraint.
To this end, as explained in section \ref{boundarytoy}, one must analyze
the induced evolution for the constraint and obtain from it
a recipe for what to provide to the incoming modes.
In the general case discussed here, notice that
$\Theta_{L\pm}$ is essentially the same that gives rise to
equation (\ref{eq:cpbc}) since is non-trivial only for the
$B_n$ and $\lambda$ components. As a result, one can follow
the same strategy and formulate constraint preserving boundary
conditions by enforcing
\begin{equation}
\dot U = UL_+ \left ( \frac{1}{2} ( F_{\phi} - F_n ) \right ) \equiv \pounds(U) \, , \label{eq:fullcpbc}
\end{equation}
where $\pounds(U)$ denotes the (maximally dissipative) constraint preserving boundary condition
defined by equation (\ref{eq:fullcpbc}).

\section{Numerical Tests}
In this section we present the results of tests aimed to
examine the solution's behavior with the different possible
choices. To this end we constructed an implementation
of the MHD equations (eqns. \ref{mainsystem}) on a 2-dimensional domain with
Cartesian coordinates $(x,y)\in[-L,L]\times[-L,L]$.
To discretize the system we employ a set of techniques which guarantee
the stability of generic linear hyperbolic systems. These techniques
are constructed to satisfy at the discrete level, the conditions holding
at the analytical one to prove well posedness of a problem through energy estimates~\cite{GKO}.
Thus we adopt discrete derivative operators satisfying summation by parts,
a 3rd order Runge-Kutta time integration and add dissipation through
a Kreiss-Oliger type operator consistent with summation by parts.
Finally, in all the tests considered we add a homogeneous `atmosphere'
throughout the computational domain in the initial data (and a `floor'
during the evolution)  for the density and the pressure to prevent negative
values from arising in the simulation. We typically adjust the atmosphere (floor) values
to be $10^{-4}~~(10^{-6})$ times the maximum of the initial values of $\rho$ and $e$ respectively.
Notice that this straightforward implementation is not expected to
handle shocks or discontinuities, however, since our goal is to examine
boundary conditions we restrict here to scenarios where these effects do
not arise within the time of interest.

For testing purposes we   adopt a few different initial scenarios which are based
on modifications of two commonly employed initial data sets: \\

\textbf{Rotor test:} This is a variation of the MHD Rotor
problem~\cite{balsaraspicer} which is smooth and allows
for considering an additional flow field and initial
constraint violation,
\begin{equation}
\rho(x,y) = \left\{
\begin{array}{l l}
  10 & \quad \mbox{if $r\leq r_0$}\\
  1+9 f(r) & \quad \mbox{if $r_0<r<r_1$ }\\
  1 & \quad \mbox{if $r_1\leq r$ } \end{array} \right.
\end{equation}

\begin{equation}
v^x(x,y) = {\nu}^x + \left\{
\begin{array}{l l}
  -f(r) v_o \frac{y}{r_0} & \quad \mbox{if $r\leq r_0$}\\
  -f(r) v_o \frac{y}{r} & \quad \mbox{if $r_0<r<r_1$ }\\
  0 & \quad \mbox{if $r_1\leq r$ } \end{array} \right.
\end{equation}

\begin{equation}
v^y(x,y) = {\nu}^y + \left\{
\begin{array}{l l}
  f(r) v_o \frac{x}{r_0} & \quad \mbox{if $r\leq r_0$}\\
  f(r) v_o \frac{x}{r} & \quad \mbox{if $r_0<r<r_1$ }\\
  0 & \quad \mbox{if $r_1\leq r$ } \end{array} \right.
\end{equation}
and
\begin{eqnarray}
p &=&1 \\
B^x &=& B_0 + {\cal B}^x \\
B^y &=& 0 \\
\phi &=& 0
\end{eqnarray}
with $r_0=1,r_1=2$, $r=\sqrt{x^2+y^2}$, $\gamma=1.4$, $v_0$ a constant and
$f(r)$ an interpolating function. For the general tests we adopt
\begin{equation}
f(r)=\frac{(r_1-r)}{(r_1-r_0)} \, .
\end{equation}
However, for the convergence test we adopt a smooth interpolating polynomial and
velocity profiles as
\begin{eqnarray}
f(r)&=& P_5(r) \\
v_o &=& \tilde v_o r^4/(1+r)^4
\end{eqnarray}
with $P_5(r)$ a polynomial of order $5$ such
that $P_5(r_0)=1; \ P_5(r_1)=0$ and both first and second
derivatives at $r=\{r_0,r_1\}$ are zero. Thus, the corresponding initial data is at least $C^2$ throughout the computational
domain.\\
Finally, the functions  $(\nu^x,\nu^y)$ and ${\cal B}^x$ are included to consider further modifications
for specific tests, in particular choosing :
\begin{itemize}
\item ${\cal B}^x = \kappa e^{-r^2}$ allows one to introduce data violating the constraints.
\item $(\nu^x,\nu^y)= -\epsilon \, \left(1-e^{-r^2}\right) \, (x,y)$ defines data inflowing to further test boundary issues.
\end{itemize}

\vspace{0.5cm}
\textbf{Blast test:} This is a variation of the MHD Blast
problem \cite{blasttest}.\\

The initial data is given by,
\begin{eqnarray}
\rho &=& 1.0 \nonumber \\
v_x &=& 0.0 \nonumber\\
v_y &=& 0.0 \nonumber\\
B_x &=& 4.0\nonumber\\
B_y &=& 0.0\nonumber\\
p &=& e^{-r^2/2}\nonumber\\
\phi &=& 0.0
\end{eqnarray}
Additionally, we also studied the solution's behavior employing
slight variations of this data set.
In particular, cases with initial incoming or outgoing velocity fields at points of the boundary.
In this way we could test different modes on the boundary which become in some cases outgoing or
incoming according to the velocities chosen.

\subsection*{Boundary conditions adopted}
For boundary conditions we adopt one of the following three possible cases:
\begin{enumerate}
\item Freezing boundary conditions. Defined by $\dot U_i = 0$, for $i=1..7$ (denoted by FR).
\item Outflow boundary condition (incoming modes set to zero). Defined as
$\Xi^+ \dot U = 0$, with $\Xi^+ = \sum_{i}  U^+_j \theta^+_j$,
for $j$ such that $\lambda_j>0$ (denoted by NI).
\item Constraint preserving boundary condition. Defined as
$\Xi^+ \dot U = \pounds(U)$, as defined by eqn. (\ref{eq:fullcpbc}) (denoted by CP).
\end{enumerate}
in all tests, and compare the solutions' behavior when employing each option. We begin
by considering the flux-conservative form of the equations (keeping $\alpha=0$) and then
examine particular cases with the Galilean invariant form ($\alpha=1$).

\subsection{Testing the implementation}
\begin{figure}
\begin{center}
\epsfig{file=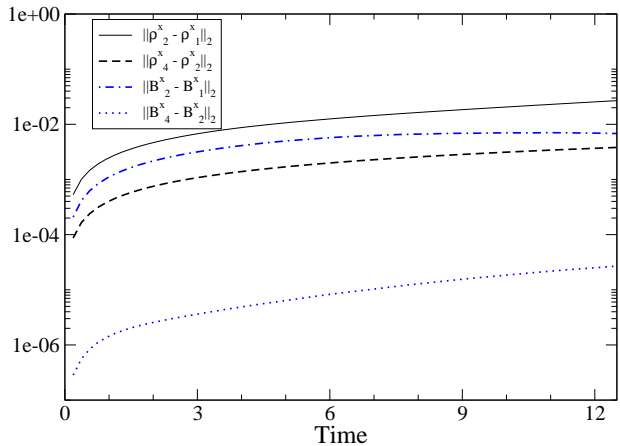,height=9.5cm,angle=270}
\caption{Behavior of the $L_2$ norm of the difference between solutions obtained with
gradually improved resolutions. The difference between them converges to zero as expected.}.
\label{fig:convergence}
\end{center}
\end{figure}

In the first test we confirm the overall convergent behavior
of the numerical solution when considering sufficiently smooth initial data.
We evolve the $C^2$ version of the Rotor initial data (with no constraint violation or
background fluid flow) for three different resolutions $\Delta_l=1.5/2^l$
($l=0,1,2$) and check the pair-wise difference of the numerical solutions obtained
decreases as expected. Figure \ref{fig:convergence} illustrates the behavior of
$||F(\Delta_l)-F(\Delta_{l+1})||_2$ (with $F$ either $\rho$ or $B^x$). As is evident
in the figure, as resolution is improved the differences decrease as expected.
For all remaining tests we present results for the finest grid employed with $l=2$.

\subsection{Blast initial data}
\begin{figure}
\begin{center}
\epsfig{file=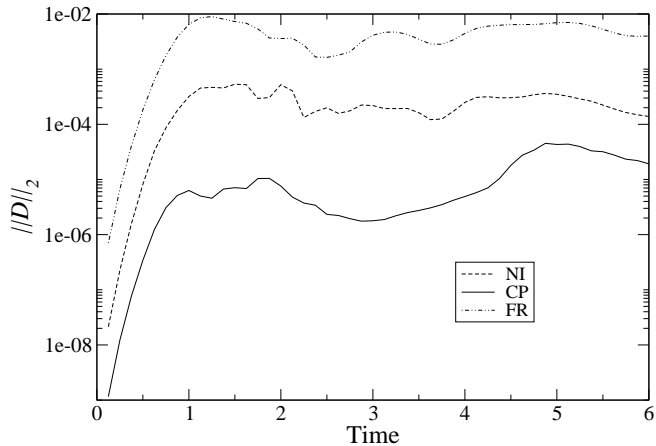,height=9.5cm,angle=270}
\caption{Behavior of the $L_2$ norm of the constraint for the different boundary
conditions. The violations in each case grow from round-off values before settling
to an approximately constant value. The values throughout the run for this norm
improves as the boundary condition adopted is refined as expected, in fact 
the solution obtained with the constraint preserving boundary condition preserves
the constraint by at least three-order of magnitude better than the one obtained
with the freezing condition.}.
\label{fig:blast}
\end{center}
\end{figure}

In this test, we adopt the blast initial data, evolve it for different choices of
boundary conditions setting $\alpha=0, c_l = 20$ and $s=1$ and examine the constraint's
behavior in each case.
Figure \ref{fig:blast} shows the $L_2$ norm of the constraint for the different
boundary value conditions.
Clearly the numerical solution obtained with constraint preserving boundary conditions
is superior by about an order
of magnitude in constraint violation than the no-incoming case and almost three
orders better than that obtained with the freezing boundary condition.
An important point to emphasize is that a closer inspection of the
solutions obtained reveals that the main contribution to the error originates at
the boundaries in all cases (though with essentially the same behavior as far
as the error's magnitude with respect to the boundary condition adopted).
The norm displayed in Figure \ref{fig:blast} is calculated over the whole computational domain
{\em ignoring} the last two points at all boundaries to avoid placing
excessive weight on the violation at the boundary. Nevertheless, as indicated,
constraint preserving boundary conditions give rise to a solution whose constraint
is violated the least.

\subsection{Rotor initial data. Effects of boundary conditions and divergence cleaning}
We adopt this data to further examine the solution's behavior and allowing
for initial violations of the constraint. To this end we adopt $\kappa = 10^{-4}$.

\begin{figure}
\begin{center}
\epsfig{file=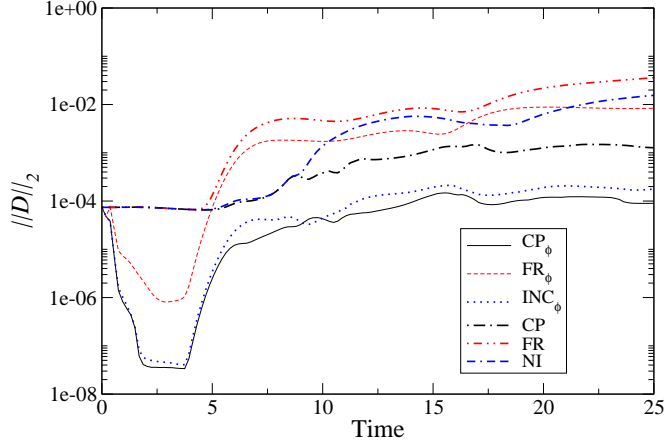,height=9.5cm,angle=270}
\caption{Behavior of the $L_2$ norm of the constraint for different options with some non-trivial
initial violation of the constraints. As is evident in the figure, the divergence cleaning
is able to damp the constraints by several orders of magnitude while the violation is
present in the bulk of the computational domain. After the solution reaches the boundary,
the boundary values induced there dominate the violation of the constraint and again the behavior
is significantly improved by the no-incoming and constraint preserving conditions.}
\label{fig:withwithout}
\end{center}
\end{figure}

Figure \ref{fig:withwithout} illustrates the solution's behavior with
the different boundary conditions and with and without the use of the divergence driver.
The addition of the divergence driver allows for a dynamical reduction
in the constraint violation until $t\simeq 4.6$, at this time the propagating
modes interact with the boundaries which become the main source of error. Here again
one sees that the constraint preserving boundary condition gives rise to significantly
smaller errors in the solution. Interestingly however, adopting the no-incoming modes
{\em together} with the constraint damping field provides a reasonably similar behavior.
This is due to the fact that the  no-incoming  boundary condition allows the
outgoing constraint violating mode to leave the computational domain, while the incoming
constraint violating mode, generated by not imposing the constraint preserving boundary
condition, is damped to a significant degree by the divergence cleaning technique.

\begin{figure}
\begin{center}
\epsfig{file=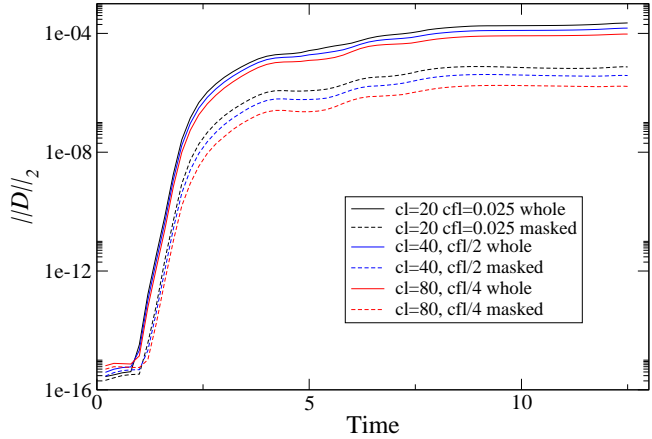,height=9.5cm,angle=270}
\caption{Behavior of the $L_2$ norm of the constraint for different values of the
coupling constant $c_l$ with constraint preserving boundary condition. The norm
is calculated over the whole computational domain (whole) or over an interior region
 of $3/4$ its size (masked). The latter illustrates how, despite not including a region
close to the boundaries, boundary induced effects are evident throughout the domain. 
As the value of $c_l$ is increased a slight improvement is achieved.}
\label{fig:rot_def_vel}
\end{center}
\end{figure}

\begin{figure}
\begin{center}
\epsfig{file=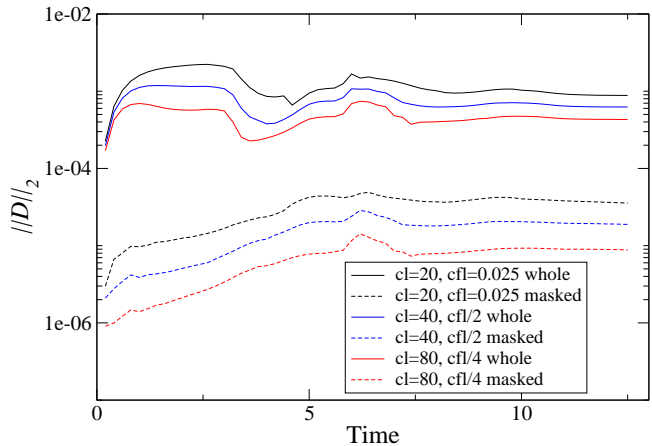,height=9.5cm,angle=270}
\caption{Similar to figure \ref{fig:rot_def_vel} but with initial data containing
constraint violations.}
\label{fig:rot_mod_vel}
\end{center}
\end{figure}

Another interesting behavior is revealed when varying $c_l$. This affects
the constraint violating modes' propagation speeds which in turn has a strong
influence in the solution's constraint behavior. In what follows we adopt the
constraint preserving boundary condition.
For this test we use the Galilean
invariant system, setting $s=0$ (so as not to include damping) and choose $c_l$ ranging 
from 10 to 80~\footnote{For smaller values
of $c_l$ ($c_l < 10 $) the code became unstable due to instabilities  generated at the corners of
the computational domain.} together with adjusting
the time step accordingly in order not to violate the Courant condition. Figures
\ref{fig:rot_def_vel} and  \ref{fig:rot_mod_vel} (with initial constraint violation)
 show the effect of varying
$c_l$ in two cases, the Rotor initial data and its modification to include
a background velocity given by $({\nu}^x=\pm 0.1x,{\nu}^y=\pm 0.1y)$.
We concentrate on
the latter case for there the effect is more striking. Notice that, in figure
\ref{fig:rot_def_vel}, the initial plateau corresponds to round-off values since the
constraint is initially satisfied to that level. Once the non-trivial part of the solution
reaches the boundary a significant violation of the constraint is generated. Depending on the boundary
condition, and on the characteristic velocities of the system under consideration that
violation propagates to the inside or just leave the domain. In our case, the ability of the
constraint preserving boundary condition to allow constraint violating modes to leave the
domain, results in smaller errors for faster propagating cases. This is a result of violations
induced by the evolution leaved the computational domain more rapidly.

\section{Conclusions}

We have investigated the numerical solution to the (ideal) Newtonian magnetohydrodynamics
equations and developed boundary conditions which at the continuum level are constraint
preserving and so would not introduce further violations in the computational domain.
At the numerical level these conditions do introduce some truncation-error level violations
which converge away with resolution. These boundary conditions developed
are consistent with maximally dissipative ones and so the discrete energy of the system
remains bounded.
We also examined the constraint's behavior when enlarging the system
so as to couple an extra field. The addition of such field, together with a suitable
modification of the equations induces a non-zero speed in the propagation
of the constraint and allows for driving its value to zero. We have studied, in particular,
two of the many equivalent systems. One which is fully conservative
--which is only weakly hyperbolic without the addition of the extra field which renders it symmetric hyperbolic--
and a Galilean invariant one which is symmetric hyperbolic --even without adding the extra field-- but
is not expressible in conservative form.

In order to examine the solution's behavior we implemented the equations in Cartesian
coordinates and with a spatial discretization that preserves the initial constraint
error. This allows us to separate bulk from boundary effects and observe that the main
violations do indeed occur at boundaries. To examine the boundary condition effects
on the evolution we adopt three types of boundary conditions:
\begin{itemize}
\item The most direct one, and
crudest, sets to zero all right hand sides in a buffer boundary region. This might lead to
inconsistencies for one might be prescribing
conditions to outgoing modes. This condition is essentially equivalent the most commonly
employed one of flux-copying near boundary points.
\item The second one is the ``no-incoming'' boundary condition where one projects out all
incoming modes in the evolution equations leaving the rest (tangential and outgoing modes)
intact. The implementation of this condition is slightly delicate as modes can change
directions over time, and so may turn from being incoming to being outgoing and vice-versa.
\item The third condition preserves the constraint in the sense that is deduced by
setting to zero the incoming mode related to the constraint propagation. This condition is
a modification to the previous one where now instead of projecting out all incoming modes, one
incoming mode is fixed so that the incoming constraint mode is zero as a result.
\end{itemize}

Not surprisingly the first option displays the worst behavior for the constraint as not only nothing
is done to minimize the constraint violations introduced through the boundaries but further
inconsistencies in the solution are induced there. As a result, in the best case the constraint 
mode bounces back from the boundary into the integration region.
The second option performed substantially better displaying a gain of an order of magnitude in
constraint preservation. This is a result of the boundary condition allowing for
one of the $\phi$ modes --carrying constraint violations-- to leave  the integration region.
With the third alternative an additional order of magnitude (at least) is gained with respect to the
non-incoming condition. This condition not only allows for constraint violations to leave
the computational domain but does not introduce significant violations at the boundary.

Thus we see that appropriately handling the boundary conditions the
problem of  constraint behavior can be significantly controlled. It should be mentioned that we
still do not have a mathematically rigorous proof that the constraint preserving
boundary conditions is well posed. However, in simpler systems --like the toy model we
discussed at the beginning of the paper-- a proof could be devised at least for
the case where the eigenvalues do not change sign. Further work in this direction is needed
to put the analysis and results obtained in stronger grounds.

\section{Acknowledgments}
We would like to thank M. Anderson, E. Hirschmann, D. Neilsen and C. Palenzuela, for
many stimulating discussions during the course of this work.
This work was supported by the National Science Foundation under grants
PHY-0326311, PHY-0554793, 0653375 to Louisiana State
University and PHY05-51164 to the Kavli Institute
for Theoretical Physics. L.L. thanks the Kavli Institute for
Theoretical Physics for hospitality where parts of this work were
completed.


\appendix
\section{Characteristic decomposition \label{chardecomp}}
In this appendix we revisit the characteristic analysis of the ideal MHD system
coupled to the divergence cleaning field (a related discussion in the absence of this
field can be found in \cite{pennisi,balsara}).
Using the decomposition on normal and tangential parts, and defining $d:=\sigma+v_n$ we get:

\begin{eqnarray}
\label{eqn:full_dec}
d \bar{\rho} \ &=& \ -\rho \bar{v_n} \, , \nonumber \\
d \bar{v}_n \ &=& \ 2 B_n/\rho \bar{B}_n - 1/\rho B_j\bar{B}^j - 1/\rho \bar{p}\, , \nonumber \\
d \bar{B}_n \ &=& \ v_n\bar{B}_n +  \bar{\phi}\, ,              \nonumber \\
d \bar{p} \ &=& \ -c_s^2\rho \bar{v}_n - (\gamma-1)B_jv^j\bar{B}_n +(1-\gamma)B_n\bar{\phi}\, ,   \nonumber \\
d B\bar{v} \ &=& \ B^2/\rho \bar{B}_n - B_n/\rho\bar{p}\, ,     \nonumber \\
d B\bar{B} \ &=& \ Bv \bar{B}_n - B^2\bar{v}_n + B_n B\bar{v} + B_n \bar{\phi}\, ,  \nonumber \\
d \tilde{\bar{v}}_i \ &=& \ \tilde{B}_i/\rho \bar{B}_n + B_n/\rho \tilde{\bar{B}}_i\, ,  \nonumber \\
d \tilde{\bar{B}}_i \ &=& \ \tilde{v}_i\bar{B}_n - \tilde{B}_i\bar{v}_n + B_n\tilde{\bar{v}}_i\, .    \nonumber \\
\sigma \bar{\phi} \ &=& \ c_l^2 \bar{B}_n
\end{eqnarray}

For the sake of clarity we will present the solution to the eigenvalue/eigenvector problem
along different cases which are naturally divided by the behavior of certain modes.

\vspace{0.5cm}

\subsubsection{BASIS}
\vspace{0.2cm}

\noindent$\bullet$ {\bf Normal modes}. These correspond to $\bar{B}_n=\bar{\phi}=0$,
where the equations become:
\begin{eqnarray}
\label{eqn:splited_normal}
d \bar{\rho} &=& -\rho \bar{v_n} \, ,  \nonumber \\
d \bar{v}_n &=&   -B\bar{B}/\rho - \bar{p}/\rho \, ,   \nonumber \\
d \bar{p} &=& -c_s^2\rho \bar{v}_n   \, ,  \nonumber  \\
d B\bar{v} &=&  -(B_n/\rho) \bar{p}  \, ,  \nonumber \\
d B\bar{B} &=& -B^2\bar{v}_n + B_n B\bar{v}   \, , \nonumber  \\
d \tilde{\bar{v}}_i &=& (B_n/\rho) \tilde{\bar{B}}_i  \, ,  \nonumber \\
d \tilde{\bar{B}}_i &=& - \tilde{B}_i\bar{v}_n + B_n\tilde{\bar{v}}_i   \, .
\end{eqnarray}
Setting $d=0$ one can show that all but the first component must vanish while itself can have any value.
So a first eigenvector is given by
\begin{equation}
U_0=(1, \ 0, \ 0, \ 0, \ 0, \ 0, \ 0),
\end{equation}
where the entries are: $U = (\bar{\rho}, \ \bar{v}_n, \
\tilde{\bar{v}}_i, \ \bar{B}_n, \ \tilde{\bar{B}}_i, \ \bar{p}, \
\bar{\phi})$.

Using all the equations we get the
eigenvalue condition:

\begin{equation}
d^4 - (c_s^2 + B^2/\rho)d^2 + c_s^2 B_n^2/\rho = 0 \, .
\end{equation}
From which we get four solutions:

\begin{eqnarray}
d_{p \pm} &=& \pm \sqrt{\left( (c_s^2+B^2/\rho) + \sqrt{(c_s^2+B^2/\rho)^2-4(B_n^2/\rho)c_s^2 } \right)/2} \\
d_{m \pm} &=& \pm \sqrt{\left( (c_s^2+B^2/\rho) -
\sqrt{(c_s^2+B^2/\rho)^2-4(B_n^2/\rho)c_s^2 } \right)/2}
\end{eqnarray}

Notice that when $B^2 \to 0$ the $d_p$ solutions tend to $\pm c_s$,
while the $d_m \to 0$ as $\|B_n\|/\sqrt{\rho}$. The $d_p$ solutions
tend to the fluid solutions while the $d_m$ to pure magnetic ones,
so they decouple in this limit. One thus must choose the
eigenvectors' normalization in a suitable way to reflect this
behavior. For the $d_p$ solutions we set $\bar{v}_n=1$, and use the
last two equations to compute the normal magnetic field and the
normal velocity, obtaining:
\begin{equation}
UP_{\pm}=\left(\mp\frac{\rho}{d_p}, \ 1, \
-\frac{B_n}{R_p\rho}\tilde{B}_i, \ 0, \ \mp \frac{d_p}{R_p}
\tilde{B}_i, \ \mp \frac{c_s^2\rho}{d_p}, \ 0\right),
\end{equation}
\noindent where $R_p:=d_{p}^2 - B_n^2/\rho$, has a finite limit as
$B^2 \to 0$.

For the $d_m$ solutions we proceed the other way around, we fix
$B\bar{B} = B_i \tilde{\bar{B}}^i=B_i M^i$, where
$M_i=\tilde{B}_i/|\tilde{B}|$ and so compute $\bar{v}_n$, and
$\bar{p}$ from the second and third equations of system
(\ref{eqn:splited_normal}), obtaining:
\begin{equation}
UM_{\pm}=\left( \frac{B\bar{B}}{R_m}, \ \frac{\mp d_m
B\bar{B}}{R_m\rho}, \ \frac{\pm B_n}{\rho \ d_m}M_i , \ 0, \ M_i , \
\frac{c_s^2 B\bar{B}}{R_m}, \ 0\right) \, ,
\end{equation}

\noindent where $R_m:=d_{m}^2-c_s^2$ has a finite limit as $B^2 \to 0$.\\

In two dimensions these are all eigenvalues-eigenvectors pairs in
the normal sector.

In three dimensions we can choose $\tilde{\bar{B}}_i=M_i$
perpendicular to $\tilde{B}_i$, but otherwise arbitrary, that is
$B\bar{B}=\tilde{B}_i \tilde{\bar{B}}^i =0$. In this case all other
components vanish except the tangential velocity which can be expressed as
$\tilde{\bar{v}}_i=d_a/B_n A_i$ with $\vec{A}=(-\tilde{B}_2,\tilde{B}_1)/|\tilde{B}|$
and $d_{a \pm} = \pm B_n/\sqrt{\rho}$ (so that the last two equations can have a
nontrivial solution). The corresponding eigenvector is:
\begin{equation}
UA_{\pm}= \left ( 0,0,\pm\frac{d_a}{B_n} A_i,0, A_i,0,0 \right ) \, .
\end{equation}

\vspace{0.5cm}

\noindent $\bullet${~\bf $\phi$ modes}\\

We now look at the modes which are induced by the introduction of the field $\phi$.
There, from the third and last equation of (\ref{eqn:full_dec}) we get a 2x2 system
 which implies: $\sigma=\pm c_l$ and so, $d := d_l = \pm c_l - v_n$.
If we fix $\bar{B}_n = b$ we can obtain the scalar components by solving
 the following subsystem of (\ref{eqn:full_dec}):

\begin{eqnarray}
\label{eqn:splited_phi}
d_l \bar{v}_n + B\bar{B}/\rho + \bar{p}/\rho &=& 2(B_n/\rho) b \, , \nonumber \\
c_s^2\rho \bar{v}_n + \bar{p}d_l &=& (-(\gamma-1)Bv + (1-\gamma)\sigma B_n)b  \, ,\nonumber \\
-B^2 \bar{v}_n -d_lB\bar{B} +B_n B\bar{v} &=& -(B_n\sigma + Bv) b \nonumber \, ,\\
(B_n/\rho)\bar{p} + d_lB\bar{v} &=& (B^2/\rho)b \, .
\end{eqnarray}
From this subsystem we get:
\begin{eqnarray}
\bar{v}_n &=& [-(d_l^2-B_n^2/\rho)F_2 + d_l^2 F_3 + d_l\rho(d_l^2 F_1 - (B_n/\rho)F_4)]/\delta_l \, , \\
\bar{p} &=& (-c_s^2 \rho \bar{v}_n + F_2)/d_l \, ,
\end{eqnarray}
where:
\begin{eqnarray}
\delta_l &=&  \rho(d_l^4 - d_l^2(c_s^2+B^2/\rho) +
c_s^2B_n^2/\rho) \, , \nonumber \\
F_1 &=& \frac{-2B_n b}{\rho}\, , \nonumber \\
F_2 &=& (1-\gamma)(Bv + B_n\sigma)b \, , \nonumber \\
F_3 &=&- F_2/(1-\gamma) \, , \nonumber \\
F_4 &=& \frac{B^2 b}{\rho} \, .\nonumber
\end{eqnarray}
Once we have $\bar{v}_n$ we can solve for the vectorial components
of the system:
\begin{eqnarray}
d_l \tilde{\bar{v}}_i - (B_n/\rho)\tilde{\bar{B}}_i &=& \tilde{B}_i b /\rho \, ,\\
d_l \tilde{\bar{B}}_i - B_n \tilde{\bar{v}}_i &=& \tilde{v}_i b -
\tilde{B}_i \bar{v}_n \, ,
\end{eqnarray}
obtaining:
\begin{eqnarray}
\tilde{\bar{v}}_i &=& [\tilde{v}_i bB_n/\rho + \tilde{B}_i(d_l b - \bar{v}_n B_n)/\rho]/\delta_s \, ,\\
\tilde{\bar{B}}_i &=& [\tilde{v}_i d_l b +\tilde{B}_i(-d_l \bar{v}_n \, ,
+ b B_n^2/\rho)]/\delta_s,
\end{eqnarray}
where $\delta_s:=d_l^2 - B_n^2/\rho$.
Combining the above intermediate steps we finally obtain,
\begin{eqnarray}
UL_{\pm} &=& \left ( -(\rho/d_l)\bar{v}_{n}, \
[-(d_l^2-B_n^2/\rho)F_2 +
d_l^2 F_3 + d_l\rho (d_l^2 F_1 - \frac{B_n F_4}{\rho}) ]/\delta_{l}, \ \right . \nonumber \\
& & \left . [\tilde{v}_i b B_n/\rho + \tilde{B}_i(d_l b - \bar{v}_n
B_n)/\rho]/\delta_s, \ b, \ [\tilde{v}_i d_l b +\tilde{B}_i(-d_l \bar{v}_n +
 \frac{bB_n^2}{\rho})]/\delta_{s}, \ \right . \nonumber \\
& & \left . \frac{(-c_s^2 \rho \bar{v}_n + F_2}{d_{l}}, \ bc_l^2/\sigma
\right )
\end{eqnarray}

\vspace{0.5cm}

\subsubsection{CO-BASIS}

\vspace{0.2cm}

Net we compute the co-basis which we will use to construct the suitable projector for enforcing
different boundary conditions. As in the previous analysis, we divide our task along different
eigenvalues for simplicity.
\vspace{0.5cm}

$\bullet$ $\Theta_{0}$ \\
Since $U_0$ has only one non-vanishing component (the first one) all elements of the co-base, except
the corresponding one have a vanishing first element.

The first element, is simply:

\begin{equation}
\Theta_0=(1, \ 0, \ 0,\ A, \ 0, \ - c_s^{-2}, \ B).
\end{equation}

To find the remaining elements we notice that if we define:
$VP:=(UL_+ + UL_-)$ and $VM:=(UL_+ - UL_-)$, the first has a zero in
the last component (corresponding to $\bar{\phi}$) while the second
has a zero in the fourth component (corresponding to $\bar{B}_n $).
Thus contraction of $\Theta_0$ with $VP$ and $VM$ will leave either
$A$ or $B$ only as unknowns. Thus we might proceed to setting $A$
and $B$  temporarily to zero, perform the contraction and extract
what these must be obtaining,
\begin{eqnarray}
A &=& \frac{-\Theta_0(VP)}{2b} \, , \\
B &=& \frac{-\Theta_0(VM)}{2bc_l} \, ,
\end{eqnarray}
where we first temporarily set to zero $A$, and $B$ in $\Theta_{0}$.
\vspace{0.5cm}

$\bullet$ $\Theta_{P\pm}$ \\
From
\begin{eqnarray}
\Theta_{P\pm}(UM_+ - UM_-)&=&0 \, \, \,  {\rm and} \nonumber \\
\Theta_{P\pm}(UM_+ +UM_-)&=&0 \nonumber
\end{eqnarray}
we get that the following structure for it:
\begin{equation}
\Theta_{P\pm} = \left(0, \ C, \ C\frac{d_{m}^2 \tilde{B}^i}{B_n
R_m}, \ D_\pm, \ \frac{-E_\pm c_s^2 \tilde{B}^i}{R_m}, \ E_\pm, \
F_\pm\right) \, .
\end{equation}
Now, from $1=\Theta_{P\pm}(UP_+ + UP_-)$ we obtain:
\begin{equation}
C=\frac{R_p\,R_m}{2(R_p\,R_m - d_m^2 \tilde{B}^2 /\rho)}\ \ \ \
\mbox{(The same for both).}
\end{equation}
While from $\pm 1 =\Theta_{P\pm}(UP_+ - UP_-)$ we deduce
\begin{equation}
E_\pm = \frac{\pm
R_p\,R_m\,d_p}{2\,c_s^2(d_p^2\tilde{B}^2-\rho\,R_p\,R_m)} \, .
\end{equation}
The remaining two components can be computed using  $VP$ and $VM$ as
above obtaining,
\begin{eqnarray}
D_{\pm} &=& \frac{-\Theta_{P\pm}({VP})}{2b} \, , \\
F_{\pm} &=& \frac{-\Theta_{P\pm}({VM})}{2bc_l}  \, ,
\end{eqnarray}
where we set to zero $D$, and $F$ in $\Theta_{P\pm}$, as in the
previous case. \vspace{0.5cm}

$\bullet$ $\Theta_{M\pm}$\\
From $ \Theta_{M\pm}(UP_+ - UP_-) = 0$ and $\Theta_{M\pm}(UP_+ +
UP_-) = 0 $ we get that the following structure for it:
\begin{equation}
\Theta_{M\pm} = \left(0, \ \frac{B_n \tilde{B}_i\,N^i_\pm}{\rho
R_p}, \ N^i_\pm, \ S_\pm, \ L^i, \ \frac{-d_p^2 \tilde{B}_i
L^i}{c_s^2 R_p \rho}, \ T_\pm\right) \, .
\end{equation}
From $1=\Theta_{M\pm}(UM_+ + UM_-)$ we obtain:
\begin{equation}
L^i=-\frac{\,R_m\,R_p\,M^i}{2(\tilde{B}^2d_p^2/\rho-R_m\,R_p)} \, .
\end{equation}


\noindent From $\pm 1=\Theta_{M\pm}(UM_+ - UM_-)$ we obtain:
\begin{equation}
N_{\pm}^i=\mp\frac{d_m\,R_m\,R_p\,M^i}{2B_n/\rho\,(\tilde{B}^2d_m^2/\rho-R_m\,R_p)}
\, .
\end{equation}



The remaining two components, $S$ and $T$ are computed as in the
previous case, using: $VP$ and $VM$ removing the $\bar{B}_n $ and
the last component $\bar{\phi}$ in $\Theta_{M \pm}$. We get,
\begin{eqnarray}
S_{\pm} &=& \frac{-\Theta_{M\pm}(VP)}{2b} \, , \\
T_{\pm} &=& \frac{-\Theta_{M\pm}(VM)}{2bc_l} \, .
\end{eqnarray}

\vspace{0.5cm}

$\bullet$ $\Theta_{A\pm}$\\
We get the co-vector of $UA$ by direct computation:
\begin{equation}
\Theta_{A\pm} = \left(0, \ 0, \ \pm \frac{B_n \, A^i}{2 \, d_a}, \
G_\pm, \ \frac{A^i}{2}, \ 0, \ H_\pm\right) \, ,
\end{equation}

\noindent where the components, $G$ and $H$ are given by:

\begin{eqnarray}
G_{\pm} &=& \frac{-\Theta_{A\pm}(VP)}{2b} \, , \\
H_{\pm} &=& \frac{-\Theta_{A\pm}(VM)}{2bc_l} \, .
\end{eqnarray}

\noindent removing the components $\bar{B}_n $ and $\bar{\phi}$ in
$\Theta_{A \pm}$.

\vspace{0.5cm}

$\bullet$ $\Theta_{L\pm}$\\
It remains now to determine the last two elements. Since the first 7
eigenvectors span completely the seven dimensional space given by
$(\bar{B}_n=\bar{\phi}=0)$ the co-vectors have only components in
that subspace and are given by:
\begin{equation}
\Theta_{L\pm} = \left(0, \ 0, \ 0, \ \frac{1}{2b}, \ 0, \ 0, \
\frac{\pm 1}{2bc_l}\right).
\end{equation}

\vspace{0.5cm}
\noindent{\bf Caveat: Singular points}\\
The characteristic structure outlined above may change as some
eigenvalues can change multiplicity for particular values of
the fields. These special cases require further analysis.
Recall the eigenvalues are given by:

\begin{eqnarray}
\sigma_0 &=& -v_n \, , \nonumber \\
\sigma_P^+ &=& d_p-v_n\, ,  \nonumber \\
\sigma_P^- &=& -d_p-v_n\, ,  \nonumber \\
\sigma_M^+ &=& d_m-v_n\, ,  \nonumber \\
\sigma_M^- &=& -d_m-v_n\, ,  \nonumber \\
\sigma_A^+ &=& d_a-v_n\, ,  \nonumber \\
\sigma_A^- &=& -d_a-v_n\, ,  \nonumber \\
\sigma_L^+ &=& c_l\, ,  \nonumber \\
\sigma_L^- &=& -c_l\, ,  \nonumber
\end{eqnarray}

\noindent they can cross when $d_p=0$, $v_n = \pm c_l$, $d_m=0$ and
$d_p=d_m$. The first two cases can be dealt with by appropriately
chosen parameters so that they would not occur in physically
relevant scenarios. For instance, the first case would imply a
vanishing sound speed and will not be considered since we will not
deal with a fluid describing dust. The second case could be avoided
by simply taking $c_l$ large enough. Therefore, we concentrate on
the other two: \vspace{0.5cm}

$\bullet$ {$d_m=0$}\\

\noindent For this we need that
\begin{equation}
\left ( c_s^2 + \frac{B^2}{\rho} \right )^2 = \left(c_s^2 + \frac{B^2}{\rho}\right)^2 -
4c_s^2 \frac{B_n^2}{\rho}
\end{equation}
which only happens when $B_n=0$.
Near that point we have,
\begin{equation}
d_m^2 = \frac{c_s^2 B_n^2}{\rho c_s^2+B^2}.
\end{equation}
 At this point we compute explicitly the eigenspace for
the coinciding eigenvalues and choose eigenvectors with good limit.
We use them in a sufficiently small neighborhood of this point using
conditional statements in the code at every point in the boundary.
\\
In three dimensions, in this case, also $d_a = 0$, therefore the
eigenvectors $UM_{\pm}$ and $UA_{\pm}$ are degenerates.

\vspace{0.5cm}

$\bullet$  $d_m=d_p$\\

\noindent For this to happen we need
\begin{equation}
 0=(d_p^2-d_m^2)^2=\left(c_s^2+\frac{B^2}{\rho}\right)^2 -
4c_s^2 B_n^2 = \left(c_s^2 - \frac{B_n^2}{\rho}\right)^2 +
2c_s^2\frac{\tilde{B}_i^2}{\rho} + \frac{\tilde{B}_i^4}{\rho^2}.
\end{equation}
Thus. $B_n^2/\rho = c_s^2$, and $\tilde{B}_i=0$. Notice that in that
case we have $R_p=R_m=0$. We proceed in a similar fashion as above.
Notice that when eigenvalues coincide all what enters in the
boundary condition is the projector on that subspace, so one just
has to choose the eigenvectors so that numerically that projector is
robust in a small neighborhood.
\\
In the case of three dimensions, $d_a^2 = B_n^2/\rho$. Hence,
$d_a=d_m=d_p$ and the eigenvectors $UP_{\pm}$, $UM_{\pm}$ and
$UA_{\pm}$ become degenerates.

\end{document}